# Properties of large-amplitudes vibrations in dynamical systems with discrete symmetry. Geometrical aspects.

Chechin George[†1], Ryabov Denis[1]
[†]e-mail:gchechin@gmail.com

[1]Institute of Physics, Southern Federal University, 194 Stachki ave., Rostov-on-Don 344090, Russia

**Abstract**

The research group from the Rostov State University has been developing the theory of bushes of nonlinear normal modes (NNMs) in Hamiltonian systems with discrete symmetry since the late 90s of the last century. Group-theoretical methods for studying large-amplitude atomic vibrations in molecular and crystal structures were developed. Each bush represents a certain collection of vibrational modes, which do not change in time despite the time evolution of these modes, and the energy of the initial excitation remains trapped in the bush. Any bush is characterized by its symmetry group, which is a subgroup of the system's symmetry group. The modes contained in the given bush are determined by symmetry-related methods and do not depend on the interatomic interactions in the considered system. The irreducible representations of the point and space groups are essentially used in the theory of the bushes of NNMs, and this theory can be considered as a generalization of the well-known Wigner classification of the small-amplitude vibrations in molecules and crystals for the case of large-amplitudes vibrations. Since using of the irreducible representations of the symmetry groups can be an obstacle to an initial familiarization with the bush theory, in the present review, we explain the basic concepts of this theory only with the aid of the ordinary normal modes, which is well known from the standard textbooks considering the theory of small atomic vibrations in mechanical systems. Our description is based on the example of describing plane nonlinear atomic vibrations of a simple square molecule.

1. **Introduction**

The present paper is a review of the theory of bushes of nonlinear normal modes (briefly "bush theory"), which represents the theory of *large-amplitude* atomic vibrations of molecule and crystal structures. Group-theoretical methods used in the theory, especially those based on the apparatus of irreducible representations of point and space symmetry groups, often create obstacles to the understanding of rather simple ideas of the bush theory.

In this review, we would like to explain the main concepts and methods of the theory of bushes of vibrational modes as simply as it possible. We also present a vast list of the original papers for deeper studying of the bush theory.

The structure of the paper can be seen from the following contents:
2. Invariant manifolds
3. Harmonic approximation and normal modes
4. Normal modes for the square molecule
5. Invariant manifolds and bushes of nonlinear normal modes for the square molecule
6. Stability of the bushes of nonlinear normal modes
7. Bushes of nonlinear normal modes in the models based on the density functional theory
8. Some mathematics
9. Conclusion

## 2. Invariant manifolds

We begin with studying the classical dynamics of $N$ material points whose interactions are described by some phenomenological potentials (model 1). This is a standard but quite rough approximation for modeling dynamics of atomic systems. Later, we will discuss model 2, which is constructed in the framework of the density functional theory, which takes into account the electron shells of atoms and their polarization during the motion of the nuclei.

In the study of dynamical systems, an important role plays the concept of *invariant manifolds*. Their search can be considered as the first step in finding exact solutions of systems of nonlinear differential equations describing a given system.

Let us explain the concept of invariant manifolds (IMs). The dynamics of a system of $N$ material points is described by a trajectory in the $6N$-dimensional phase space of all their coordinates and momenta (velocities). By the definition, if the system's initial state, determined by the coordinates and velocities of all its particles, lies on a given IM, then the entire phase trajectory will lie on this manifold, i.e. the system does not leave this manifold at any time during its evolution. Obviously, any exact solution of the dynamical equations must lie on a certain IM, and this fact determines the importance of finding such manifolds.

Unfortunately, in mathematics, there are no general methods for searching invariant manifolds for arbitrary dynamical systems. On the other hand, if the system has some *discrete symmetry*, then one can propose general methods for constructing such manifolds. The key idea is that invariant manifolds correspond to the *subgroups* of the system's symmetry group. The theory of bushes of nonlinear normal modes starts with this idea [1-3], and we will explain it with the aid of our model 1.

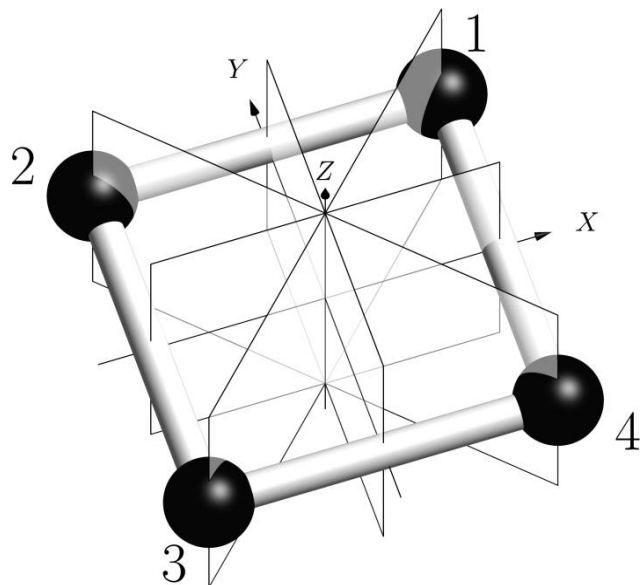

Fig. 1. The model of a simple square molecule.

In Figure 1, we show a planar square molecule consisting of four atoms located at the corners of the square. We consider this molecule in space and, therefore, one can think about it as the square base of the Cheops pyramid with point symmetry group $C_{4v}$ in the Schoenflis notation. This group consists of 8 elements. They are:

– vertical fourth-order axis, which determines rotations by 0, 90, 180 and 270 degrees around the $Z$ axis (the formal "rotation" by 0 degrees is just the identity element of the group);

– four reflection planes that pass through the vertical axis of the pyramid. Two of them are "coordinate planes"—they pass through the axes $X$ and $Y$, respectively, while the other two reflection planes are "diagonal"—they pass through the diagonals of the square.

### 3. Harmonic approximation and normal modes

The standard approach for studying atomic vibrations with small amplitudes in molecules and crystals is the *harmonic approximation* [4]. In this approximation, the potential energy of the nonlinear system is expanded into the multidimensional Taylor series over all displacements of atoms from their equilibrium positions and all members of this series, starting with cubic ones, are discarded. As a result, one obtain a quadratic form (there are no linear terms since decomposition is done in the vicinity of the equilibrium state). Then the classical Newton equations, determined by such quadratic potential energy, form a system of *linear* differential equations with constant coefficients with respect to the atomic displacements ($x_j$).

This system can be split into *independent* equations if the above quadratic form is reduced to the canonical form, in which it takes the form of a superposition of only the *squares* of the variables $x_i^2$ (all terms of the type $x_i x_k$ are eliminated).

This can be achieved by using a linear orthogonal transformation of the force constant matrix (the matrix of the second partial derivatives of the potential energy with respect to all variables) to the *diagonal* form. As it is well known, this problem is reduced to finding all the eigenvalues and eigenvectors of the force constant matrix by an appropriate orthogonal transformation in the space of old variables.

The new dynamical variables $y_j(t)$, in contrast to the old variables $x_i(t)$, are nothing more than *normal modes* of the original dynamical system.

These modes are independent of each other in the harmonic approximation, and their number is equal to the number of the system's degrees of freedom.

If some *weak nonlinear terms* in potential energy are taken into account, the normal modes begin to interact with each other, and their interactions can be calculated in the framework of the perturbation theory. Let us note that the independence of normal modes from each other in the harmonic approximation allows one to use the quantum-mechanical description of atomic vibrations in crystals by introducing the concept of phonons, which are quantized normal modes. Similarly, other quasi-particles (magnons, etc.) are introduced in crystal physics by quantizing normal modes of different physical nature.

### 4. Normal modes for the square molecule

Let us consider all normal modes for the case of atomic vibrations *in the plane* of the square molecule presented in Fig. 1. This molecule possesses 8 degrees of freedom because each atom has two degrees of freedom (displacements along the *X* and *Y* axes) and, therefore, 8 normal modes. A certain pattern of atomic instantaneous displacements, which are shown by arrows in Fig. 2, corresponds to each of the normal modes.

All these normal modes were obtained for the case when atoms interact via the Lennard-Jones potential in "standard form" (both phenomenological constants corresponding to the contribution from the attraction energy and repulsion energy are set equal to unity). The eigenvalues of the force constant matrix, i.e. squares of the normal modes frequencies are $\lambda_1=32.575$, $\lambda_2=\lambda_7=\lambda_8=0$, $\lambda_3=34.541$, $\lambda_4=-1.965$, $\lambda_5=\lambda_6=33.986$. In this figure, modes $\varphi_7$ and $\varphi_8$ are not indicated. They describe the motion of the molecule as a whole along the *X* and *Y* axes (zero eigenvalues correspond to them).

The atomic displacement patterns corresponding to the normal modes, which are given in Fig. 2, were obtained from the eigenvectors of the force constant matrix.

Let us note that the form of the atomic patterns (the geometrical aspect of the problem) is completely *independent* of the interparticle interactions. However, vibrational frequencies (the dynamic aspect of the problem) depend substantially on the interatomic interactions and we discuss this question below.

We see that the atomic patterns in Fig. 2 for all normal modes have some symmetry, and the corresponding point groups in the Schoenflies notation are also shown in Fig. 2. Let us consider the form of these patterns in more detail.

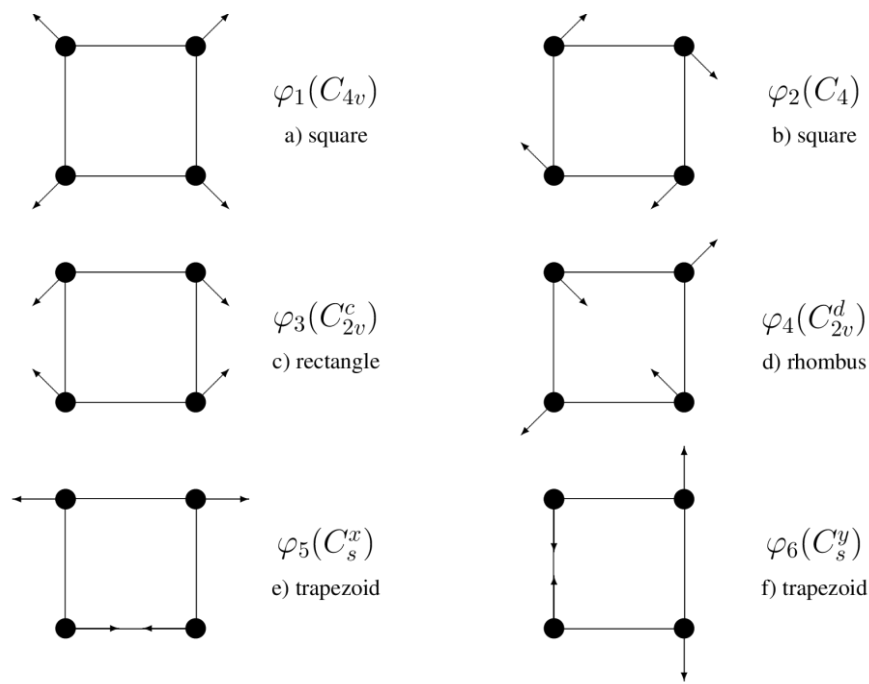

Fig. 2. Normal modes for the square molecule.

First of all, it should be emphasized that the lengths of all arrows, showing atomic displacements, are of the *same length* for any chosen mode!

Figure 2a shows the instantaneous deformation of molecule corresponding to the so-called "breathing mode" $\varphi_1$: molecule retains its *square shape*, which periodically increases and decreases in size. The mode in Fig. 2b describes the rotation of our square molecule as a whole around its center according to mode $\varphi_2$. Fig. 2c shows the rectangular deformation of the molecule during its oscillations according to mode $\varphi_3$: it periodically stretches along one coordinate axis, contracting along the second axis, and vice versa.

Figure 2d shows the rhombic deformation of the molecule during its oscillations in mode $\varphi_4$—it either stretches along one diagonal, contracting along the second diagonal, and vice versa.

Finally, the trapezoidal deformation of our molecule with a different orientation along the coordinate axes is depicted in Figs. 2e and 2g. It corresponds to the modes $\varphi_5$ and $\varphi_6$.

Atomic patterns corresponding to all normal modes of the square molecule, which are depicted in Fig. 2, were obtained as eigenvectors of the force constant matrix for the case of Lennard-Jones potential. However, it is important to emphasize that the same types of patterns can be obtained with the aid of *basis vectors* of all irreducible representations of the molecule's symmetry group *without taking into account* any specific interatomic potentials.

It is noteworthy that eigenvalue, corresponding to the rhombic normal mode $\varphi_4$, turns out to be a *negative* number and, therefore, the frequency of "oscillations" of this mode is imaginary. This means the *instability* of the square molecule relative to the rhombic deformation, i.e. the molecule does not oscillate with respect to the square equilibrium positions but leaves them.[1]

Starting the study of normal modes in a square molecule, we implicitly assumed that its square configuration is stable and small atomic vibrations occur relative to this configuration. However, the results of our calculations for the case when atoms interact via the Lennard-Jones potential show that the square configuration of the molecule, in this case, is unstable.

---

[1] It may seem that as a result, the molecule will come to an equilibrium configuration in the form of the rhombus with an angle of 60 degrees. Oddly enough, but the equilibrium rhombus has a slightly larger acute angle 60.23°. But with the formation of a plane grid of many atoms, this angle gradually tends to 60 degrees as the size of the grid increases – giving rise to a plane hexagonal lattice.

It does not follow from this fact that the square configuration is unstable at other interatomic interactions. For example, in the case of Morse interactions, the parameters of the potential can be chosen in such a way that the square molecule will be stable. Even in the case of the Lennard-Jones potential, one more atom can be placed in the *center* of the square with a sufficiently large coefficient, corresponding to the attraction energy, in order to achieve stability of such a centered molecule. Let us note, in this regard, that precisely centered plane square molecules (for example, the molecule $XeF_4$), exist in nature.

In any case, our group-theoretical analysis does not depend on the centering of the square. Indeed, we exclude modes corresponding to the movement of the molecule along coordinate axes, assuming thereby the immobility of the molecule center where the additional atom can be placed.

## 5. Invariant manifolds and bushes of nonlinear normal modes for the square molecule

Here we discuss invariant manifolds that can be obtained with the aid of group-theoretical methods. Since their existence is dictated by the molecule symmetry, we call them "symmetry-determined invariant manifolds".

In the framework of classical dynamics, which is described by Newton's differential equations, the dynamical behavior of our molecule is uniquely determined by its initial state. This is a situation of so-called *classical determinism*—"the past completely determines the future".

In such a situation, we can say that there exists a certain "law of conservation of symmetry". This means that if some element of symmetry of the system's motion exists at the initial moment (it is determined by the atomic displacement pattern), then this element exists during the whole time of movement until it is stable. On the other hand, at the point of stability loss, the symmetry of the system may decrease similarly to the spontaneous symmetry decrease occurring in the case of continuous structural phase transitions in crystals (Newtonian mechanics uniquely determines the movement before and after the point of the stability loss, but not at this point itself).

Let us consider the atomic displacement pattern corresponding to the mode $\varphi_4$, which represents a rhombus with diagonals of the *same length* since, as already was mentioned, all arrows showing displacements of the atom from the corners of the square, possess the same length in the framework of the harmonic approximation. The symmetry group of this pattern is $C_{2v}^d$. However, the considered pattern is not one of the most general form, which corresponds to the point group $C_{2v}^d$! Obviously, the rhombus with diagonals of *different lengths* also has the same symmetry group and it corresponds to displacements of different lengths for atoms located on different diagonals.

Thus, the invariant manifold corresponding to the point group $C_{2v}^d$ is wider than that which we obtained for small oscillations in the harmonic approximation. From this point, the introduction to the bushes of vibrational modes starts.

### 5.1 Bushes of vibrational modes

Displacements of atoms from their equilibrium positions in the plane of the molecule can be described by a *configuration vector* of the form:

$$X = [x_1, x_2 / x_3, x_4 / x_5, x_6 / x_7, x_8] \qquad (1)$$

where the first two coordinates are the *x*- and *y*- displacements of the first atom, the next two coordinates are *x*- and *y*- displacements of the second atom, etc. (the atom's numbers are shown in Fig. 1).

In these notations, the eigenvector of the force constant matrix, corresponding to the $\varphi_4$ mode and describing the rhombus of its atomic pattern, has the following form:

$$X = [a, a / a, -a / -a, -a / -a, a] = a\,[1, 1 / 1, -1 / -1, -1 / -1, 1].$$

On the other hand, the coordinate vector corresponding to the rhombic pattern of the *general form* with the symmetry group $C_{2v}^d$ is

$$X_4 = [a,a\ /\ b,-b\ /\ -a,-a\ /\ -b,b]. \tag{2}$$

This vector determines the rhombus with diagonals of *different lengths*. If we decompose it over the complete set of eigenvectors of the force constant matrix (note that by virtue of Wigner's theorem we can look at them as basis vectors of irreducible representations of the symmetry group $C_{4v}$), then it is easy to verify that there are *only two* nonzero terms in such decomposition and they correspond to the modes $\varphi_4$ and $\varphi_1$:

$$X_4 = A\ [1,1\ /\ 1,-1\ /\ -1,-1\ /\ -1,1] +$$
$$+ B\ [1,1\ /\ -1,1\ /\ -1,-1\ /\ 1,-1], \tag{3}$$
$$A = (a+b)/2, \quad B = (a-b)/2.$$

If we choose initial conditions for solving nonlinear dynamical equations of our molecule in the form of (2) for some fixed values of $a$, $b$ [or $A$, $B$ in (3)], we find that these constants begin to change during the numerical integration of these equations, i.e. they are certain *functions of time*, $a(t)$ and $b(t)$. The configuration vector $X_4(t)$ will possess the form (2) for any instant $t$, i.e. it does not leave the invariant manifold determined by Eqs. (2) or (3):

$$X_4(t) = A(t)\ [1,1\ /\ 1,-1\ /\ -1,-1\ /\ -1,1] +$$
$$+ B(t)\ [1,1\ /\ -1,1\ /\ -1,-1\ /\ 1,-1]. \tag{4}$$

Thus, we have obtained *two-dimensional* bush B4 with the symmetry group $C_{2v}^d$.

On the other hand, for vibrations corresponding to the symmetry group $C_{4v}$, starting from the normal mode $\varphi_1$, we will obtain the dynamical regime of the form

$$X_1(t) = A(t)\ [1,1\ /\ 1,-1\ /\ -1,-1\ /\ -1,1]. \tag{5}$$

This is the *one-dimensional* bush with $C_{4v}$ point group.

Using the above examples, we can introduce some notions and present some ideas of the theory of bushes of nonlinear normal modes.

– We have excited bush B4 with the aid of excitation of normal mode $\varphi_4$. If its amplitude is sufficiently small, we will see during integrations of nonlinear equations practically only this mode in full accordance with the theory of small atomic vibrations. However, as the initial amplitude of $\varphi_4(t)$ increases, the additional mode $\varphi_1(t)$ can be observed in the numerical solution of nonlinear equations since nonlinear terms begin to play a more and more important role. We call the primarily excited mode $\varphi_4(t)$ the "root mode" of the bush, while the mode $\varphi_1(t)$, involving into the vibration process because of its interaction with the root mode, we call the "secondary mode".

– It is important to note that our secondary mode $\varphi_1(t)$ possesses a higher symmetry group ($C_{4v}$) with respect to the group $C_{2v}$ of the root mode. It can be proved in the framework of the bush theory that this is a general property—symmetry of all secondary modes always *equal or higher* than that of the root mode. Note that the above-discussed notions are similar to primary and secondary order parameters in the theory of phase transitions in crystals [5].

– One should not think that secondary modes always are smaller in amplitude than the root mode—they can be of the same order of magnitude for the case of large nonlinearity.

– A given bush can be certainly excited by simultaneous excitations of all its modes, not only by one root mode.

– One-dimensional bush represents a *periodic* motion, while the $m$-dimensional bush represents a *quasi-periodic* motion with $m$ main frequencies (and their different integer linear combinations) in its Fourier spectrum.

Let us now consider bush B2 generated by the root mode $\varphi_2(t)$, that describes the rotation of our square molecule as a whole (see its atomic pattern in Fig. 2a).

Being conventional normal mode, it is independent of all other modes in the harmonic approximation. But if we take into account nonlinear terms in the decomposition of the molecule Lennard-Jones potential energy, the mode $\varphi_2(t)$ involves into the vibrational process one more mode $\varphi_1(t)$ (and only this mode!). This is the same secondary mode as that in the above-considered bush B4. As a result, the connection between rotation and oscillation movements appears, i.e. our molecule rotates and simultaneously oscillates. The connection between these two types of motion is a well-known property of real molecules.

The complete list of all bushes, which can be excited in the square molecule, is presented in the Table 1. Each of them is determined by a certain subgroup of the molecule symmetry group $C_{4v}$. Therefore, to find all bushes we must consider all such subgroups, taking into account their *different settings* into the parent group $C_{4v}$. Each subgroup determines the certain invariant manifold, which then we must decompose into the basis vectors of the irreducible representations of the parent group to single out all modes forming the bush. It is obvious from the described procedure that bushes found in this way fully *independent* of any specific interatomic interactions and these group-theoretical results are exact. This is a geometrical aspect of the bush theory.

Table 1. Vibrational bushes for a square molecule.

| Number of the bush | Irreps which contribute to a given bush | Bush | Symmetry group $G \subseteq G_0$ |
|---|---|---|---|
| 1 | $\Gamma_1$ | $\mu_1 \varphi_1$ | $C_{4v}$ |
| 2 | $\Gamma_1; \Gamma_2$ | $\mu_1 \varphi_1 + \mu_2 \varphi_2$ | $C_4$ |
| 3 | $\Gamma_1; \Gamma_3$ | $\mu_1 \varphi_1 + \mu_3 \varphi_3$ | $C_{2v}^c$ |
| 4 | $\Gamma_1; \Gamma_4$ | $\mu_1 \varphi_1 + \mu_4 \varphi_4$ | $C_{2v}^d$ |
| 5 | $\Gamma_1; \Gamma_3; \Gamma_5$ | $\mu_1 \varphi_1 + \mu_3 \varphi_3 + \mu_5 \varphi_5$ | $C_s^c$ |
| 6 | $\Gamma_1; \Gamma_4; \Gamma_5$ | $\mu_1 \varphi_1 + \mu_4 \varphi_4 + \mu_5 (\varphi_5 + \varphi_6)$ | $C_s^d$ |
| 7 | $\Gamma_1; \Gamma_2; \Gamma_3; \Gamma_4$ | $\mu_1 \varphi_1 + \mu_2 \varphi_2 + \mu_3 \varphi_3 + \mu_4 \varphi_4$ | $C_2$ |
| 8 | $\Gamma_1; \Gamma_2; \Gamma_3; \Gamma_4; \Gamma_5$ | $\mu_1 \varphi_1 + \mu_2 \varphi_2 + \mu_3 \varphi_3 + \mu_4 \varphi_4 + \mu_5 \varphi_5 + \mu_6 \varphi_6$ | $C_1$ |

Each bush in Table 1 is presented as a linear combination of certain modes. The second column shows the symbols of irreducible representations (irreps) of the symmetry group $C_{4v}$, which contribute to a given bush. There are five irreps in this group—four one-dimensional representations ($\Gamma_1; \Gamma_2; \Gamma_3; \Gamma_4$) and one two-dimensional representation ($\Gamma_5$). At this stage of the description of the bush theory, this column can be ignored—we will return to the discussion of connection of bushes with irreducible representations of symmetry groups in Section 8 of the present paper.

Above, we have briefly described some of the geometrical aspects of the bush theory, and now we would like to discuss several important issues related to their dynamics.

**5.2 Rosenberg nonlinear normal modes in the framework of the bush theory**

Now we should like to explain, why we use the term "bushes of *nonlinear normal* modes" (NNMs).

In 1962 Rosenberg introduced the concept of the "similar nonlinear normal mode" [6]. By definition, this is a dynamical regime in $N$-particle mechanical system for which the evolution of all degrees of freedom is described by one and the same function of time $f(t)$:

$X(t) = [a_1, a_2, \ldots, a_N] f(t),$ (6)

where $a_i$ ($i=1..N$) are constant coefficients determining the amplitudes of vibrations of all the degrees of freedom.

Conventional (linear) normal modes also satisfy the definition (1) with $f(t) = \sin(\omega t + \varphi_0)$. In the general case, the function $f(t)$ for the Rosenberg mode is determined by a certain differential equation (the so-called "governing" equation).

Unlike the case of the conventional normal modes, the number of nonlinear normal modes does not have any relation to the dimension of the considered dynamical system (as a rule, this number is lesser than the system's dimension.)[2]. As a consequence, NNMs cannot be used as a basis of the configuration space.

It is very important to note that Rosenberg NNMs can exist only in the systems with very specific properties, for example in those whose potential energy represents a *homogeneous function* of all its variables (obviously, this is the very exotic case).

However, one can prove in the framework of the bush theory that in the systems with discrete symmetry there can exist Rosenberg NNMs fully independent of the type of interatomic interactions. The existence of these modes is dictated only by symmetry-related properties, because of which we call them "symmetry-determined NNMs". Only such Rosenberg modes are considered in this paper.

Comparison of Eq. (5) and Eq. (6) shows that any one-dimensional bush is a Rosenberg mode. On the other hand, we can say that the two-dimensional bush B4, described by Eq. (3), contains *two different virtual* Rosenberg modes—the root mode

$$A(t) [1,1 / 1,-1 / -1,-1 / -1,1] \quad (7)$$

and the secondary mode

$$B(t) [1,1 / -1, 1 / -1,-1 / 1,-1]. \quad (8)$$

Why do we call these modes of the bush B4 by the term *virtual* Rosenberg modes? The matter of the fact is that the mode (7) cannot exist without the mode (8) (the last mode automatically involves into the vibrational process by the former mode), while the mode (7) can exist as an independent Rosenberg mode only in the case when this mode will be excited at the initial time without excitation of any other vibrational modes.

Since any vibrational regime in a system with the symmetry group $G$ can be associated with a certain subgroup of this group, while some bush corresponds to every subgroup, we can refer to any vibrational regime in the dynamical system with discrete symmetry as to the *bush of nonlinear normal modes*.

### 5.3 Parent symmetry group

The parent group is the group characterizing the symmetry of the considered system. Different physical ideas may be used for choosing the parent group. Wigner in his classical paper on classification of normal modes in systems with discrete symmetry [7] chose as parent group the symmetry group of the system's *equilibrium state*. However, most often for this role the symmetry group of the system's *Hamiltonian* is chosen. We would like to explain what is the difference between these groups. The symmetry group of Hamiltonian can be higher than that of the equilibrium state. Indeed, this group can dictate the existence of several *symmetry-equivalent* equilibrium states that transform into each other due to the action of some of its elements. For example, this is a typical situation in perovskite $BaTiO_3$, where around titanium atom at the center of the cubic cell there are several symmetrically located local energy minima, each of which has less symmetry compared to their complete ensemble. During the phase transition of the displacement type, the titanium atom passes into one of these minimums, as a result of which the symmetry of the system decreases. In the case of the ordering type transition, the titanium atom can be found at different minima with different probabilities.

What will be the difference in the classification of bushes with a different choice of the parent group? If certain subgroups of the initially chosen parent group were obtained and then we find that the system has some *additional* symmetry elements, i.e. the parent group is higher, then the set of the initially found bushes *is changed* because of combining some old bushes into one larger bush with a higher own symmetry, as well as of appearing some new bushes. For example, in arbitrary monoatomic chains, there can exist only three one-dimensional vibrational bushes, but if the potential of interatomic interactions is symmetric with respect to each site of the chain, then there can exist already five such bushes, depending on the number of atoms in the chain [8-11].

---

[2] In some very special cases, the number of NNMs can be even greater than this dimension.

If the dynamical system is not Hamiltonian, then the symmetry group of its equations can be taken as the parent group.

To check, which choice of the parent group is the most appropriate for a given system, one can perform its modeling using methods of molecular dynamics or density functional theory methods. Based on the results of such modeling, it is necessary to reveal which secondary modes and with what amplitudes are involved into the vibrational process when the root mode of the bush is excited.

Note that the root mode is the mode of the bush that has minimal symmetry among all other its modes.[3]

### 6. Stability of the bushes of nonlinear normal modes

As was already discussed, the phase trajectory of the considered system does not leave a given symmetry-determined invariant manifold during its time evolution. This conclusion is based on the classical determinism provided by the uniqueness of the solution of Newton equations for the given initial conditions. If this solution does not practically change by any sufficiently small perturbations, which always present in any physical system, one can speak about the case of *stability by Lyapunov* [12].

However, in the case when such stability is lost, even infinitely small fluctuations can lead to fully unpredicted results. For example, let us refer to the phenomenon of *parametric resonance* of the vertically hanged pendulum whose length is periodically changed in time with the frequency $\Omega$ and amplitude $\varepsilon$. There are no any external horizontal forces acting on the pendulum in the Newton equations. However, for the case $\Omega=2\omega$, where $\omega$ is eigenfrequency for the mean length of the pendulum, it begins to oscillate in the horizontal direction even in the case of infinitely small $\varepsilon$.

Investigation of the above parametric resonance in pendulum motion can be reduced to studying the stability of zero solution of the Mathieu equation. For this equation, there is a well-known Ince-Strutt diagram from which one can see the infinite set of stable and unstable zones in the amplitude-frequency plane.

As we have already seen, the root and secondary modes, containing in a given bush, are active modes. All other vibrational modes are "sleeping" with respect to this bush since they possess zero amplitudes. When the amplitude of the root mode increases, its frequency changes (this is the main property of nonlinear oscillations!). At some value of the root mode amplitude, the so-called "critical" amplitude, a parametric resonance can occur between the root mode frequency and the frequency of one of the sleeping modes.

As a result, this sleeping mode wakes up, i.e. becomes active. Thus, the number of active modes in the system *increases*, and a new bush of a *larger dimension* and a *lower symmetry* arises.

The critical amplitude of the root mode substantially depends on the specific interatomic interactions and can be investigated in the framework of the dynamical part of the bush theory. We would like to note that there is a deep connection between the geometrical and dynamical aspects of the theory of the bushes of nonlinear normal modes, which was studied in details in [3]. The dynamic aspects of this theory will be discussed in the second part of the present review.

### 7. Bushes of nonlinear normal modes in the models based on the density functional theory

Our second model is the model based on the density functional theory (DFT). Quantum-mechanical models in the framework of this theory are much more adequate to reality [13]. Indeed, the typical accuracy of such models in atomic coordinates is about 1% and about 10% in binding energy. They take into account quantum characteristics of atoms and additional degrees of freedom corresponding to the electrons of atomic shells. The well-known software packages for treating DFT-models, such as ABINIT [14], Quantum ESPRESSO [15], VASP [16], allow one to consider dynamical problems in the Born-Oppenheimer approximation, and to take into account the influence of atomic-shell polarization on the classical motion of nuclei.

---
[3] Here, we do not consider the more complicated case when the root mode corresponds to the intersection of several symmetry groups.

Since the density functional theory is based on the Schrödinger multielectron equation, which is a *linear* equation, one can ask what new information above that of the Wigner's classification of atomic vibrations by individual irreducible representations of the system's symmetry group can be obtained from the theory of the bushes of nonlinear normal modes?

The matter is that despite of the linearity of the exact Schrödinger equation, the well-known approximate methods for solving this equation, such as Hatree-Fock method and Kohn-Sham method of the density functional theory are based on the solving of some systems of *nonlinear integro-differential equations*.

Therefore, all group-theoretical methods of the bush theory can be used for studying large-amplitude atomic vibrations in molecular and crystal models in the framework of DFT-theory. Such studies were performed in [17-20]. In these works, the transfer of excitation between modes of different symmetries is investigated, and the selection rules for such a transfer based on the bush theory were confirmed. On the one hand, these works verify the applicability of the bush theory to DFT-models and, on the other hand, check the correctness of our cumbersome group-theoretical calculations of this theory.

## 8. Some mathematics

The key idea of the bush theory is very simple—we must find all symmetry-determined invariant manifolds from the condition of the configuration vector invariance with respect to each subgroup of the system's parent symmetry group. However, the technical realization of this idea is rather complicated, and below we very briefly and schematically outline the main points of the group-theoretical approach for studying bushes of nonlinear modes.

At the end of the 70s of the last century, we developed a set of computer programs for the group-theoretical analysis of the so called "complete condensate" of the primary and secondary order parameters corresponding to the structural phase transitions in crystals [21]. Later, this complex was generalized for studying bushes of nonlinear normal modes. Below, we consider some details of the group-theoretical methods of the bush theory.

1. Let us begin with the case when we already know a certain subgroup $G_j$ of the parent group $G$. The invariant manifold, corresponding to this subgroup, can be found from the condition of configuration vector invariance:

$$\hat{G}_j X_j = X_j. \qquad (9)$$

Here $X_j$ is the invariant vector, while $\hat{G}_j$ is the group of operators, isomorphic to the symmetry group $G_j$, which act in multidimensional vector space of all $N$ degrees of freedom (note that elements of the group $G_j$ act in ordinary three-dimensional space).

On the other hand, the vector $X_j$ can be decomposed into basis vectors $\varphi_i$ of all irreducible representations (irreps) $\Gamma_i$ of the parent space group $G$:

$$X_j = \Sigma_i\, C_{ji} \varphi_i \qquad (10)$$

The invariance condition (9) corresponds to the whole $N$-dimensional space. The similar invariance conditions for all the individual irreps $\Gamma_i$ can be obtained from it:

$$(\Gamma_i \downarrow G_j)\, C_{ji} = C_{ji} \qquad (11)$$

Here $\Gamma_i \downarrow G_j$ is the so-called restriction of the irrep $\Gamma_i$ on the subgroup $G_j$, which is the set of matrices of the representation $\Gamma_i$ corresponding to all elements of the group $G_j$.

For every $n_i$-dimensional irrep $\Gamma_i$, Eq. (11) represents a certain system of $n_i$ *linear* algebraic equations whose general solution depends on several arbitrary constants ($a$, $b$, $c$, etc.). If this solution is zero, irrep $\Gamma_i$ does *not contribute* to the bush with the symmetry group $G_j$.

The vector $C_{ji}$ being multiplied by the basis vectors of the irrep $\Gamma_i$ (see about these vectors below) determines a concrete form of the vibrational mode $\varphi_j$ associated with this irrep, i.e. the pattern of atomic displacements corresponding to the obtained mode.

According to Eq. (11), the vector $C_{ji}$ is conserved by all matrices of the irrep $\Gamma_i$, which correspond to elements of the group $G_j$. However, *some other matrices* of this irrep, corresponding to elements of the parent group $G$, which are not contained in its subgroup $G_j$, can also conserve the vector $C_{ji}$. Therefore, this vector can select some additional symmetry elements of the group $G$ (except those that are already included in the subgroup $G_j$), i.e. it corresponds to a subgroup $\tilde{G}_j$ of a *higher order* than $G_j$.

The subgroup $\tilde{G}_j$ of minimal symmetry represents the symmetry group of the whole bush B$j$ and, at the same time, the symmetry group of its root mode.[4]

Equation (11) is the source of the *selection rules* for excitation transfer from the root mode to the secondary ones.

At this point, it is appropriate to explain the words "the mode belonging to a given *multidimensional* irrep". Let us obtained a certain invariant vector $C_{ji}$ as the general solution of Eq. (11) for a six-dimensional irrep $\Gamma_i$, which has the form $[a,b,-a,c,-b,c]$.[5] Then the vibrational mode, corresponding to this vector, is

$$X(t) = a(t)\varphi_1 + b(t)\varphi_2 - a(t)\varphi_3 + c(t)\varphi_4 - b(t)\varphi_5 + c(t)\varphi_6,$$

where $\varphi_k$ ($k$=1..6) are *basis vectors* of the representation $\Gamma_i$.

The complete set of the bush modes with symmetry group $G_j$ is determined by the collection of all nonzero solutions of the equations (11).

2. In order to construct an explicit form of atomic displacement patterns, it is necessary to obtain basis vectors of those irreps of the parent group $G$, which are contained in the decomposition of the reducible *mechanical representation* of this group into irreducible representations. Usually, such basis vectors are obtained by the projection operators technique. In our paper [21], a more transparent and more convenient for our purposes method was presented. This is the so-called "direct method" based on the definition of the group representation.

If the invariant vector, corresponding to the subgroup $G_j$, and basis vectors of the irrep $\Gamma_i$ are already found, then the contribution of this representation to the atomic displacement pattern can be obtained by the Eq. (10).

3. A very difficult problem of the bush theory is finding of all subgroups $G_j$ of the parent space group $G$, taking into account all their possible settings in the group $G$. For this purpose, we developed a special group-theoretic algorithm based on the sequential finding of all possible pair-wise intersections of invariant vectors of a given representation[6],[7].

4. A separate important and cumbersome problem is obtaining the so-called "full" irreducible representations of the space groups. For this purpose, we used in our computer program the standard algorithm based on the method of projective irreducible representations, which is well described in [22]. Let us note that it is very difficult to find the full irreps of the space groups using tables from the book by Kovalev [23], due to the deep degree of embedding notations of different levels in each other. For solving specific problems associated with the construction of bushes of modes for structures with space symmetry, we use our own tables in the so-called "genesis" form[8].

### 9. Conclusion

In 1930, Wigner published his pioneering work on the application of group-theoretical methods for studying and classification of small oscillations of systems with discrete symmetry [7]. This approach has become classical and has been included in all textbooks on the physics of molecules and crystals. Wigner showed that small vibrations within the framework of the *harmonic approximation* can be classified by *irreducible representations* (irreps) of the symmetry groups of these objects. Indeed, the degeneracy degree of a given fundamental frequency is equal to the *dimension of a certain* irreducible representation of the symmetry group, while the corresponding normal modes are "transformed according to this representation", i.e. they can be considered as basis vectors of this irrep. Wigner's theorem [24] determines the general form of the Hamiltonian matrix corresponding to small oscillations of the system with discrete symmetry.

---

[4] We omit here some subtleties of the discussed problem.
[5] Six-dimensional irreps frequently occur in the cubic space groups and a few dozens of different invariant vectors with different own symmetry groups $G_k$ can be associated with each of these irreps.
[6] Each of them represents some set in the space of the given irrep.
[7] We used this algorithm earlier to construct a complete condensate of order parameters in the theory of structural phase transitions in crystals [5].
[8] For many space groups, we published them earlier as VINITI deposited manuscripts.

We emphasize once again that all these results relate only to *small* atomic oscillations in the framework of the harmonic approximation. In contrast, the bush theory presents a group-theoretical classification and corresponding methods for studying dynamical regimes in the systems with discrete symmetry for *any amplitudes* of the atomic vibrations.

In this review, we have tried to explain the basic geometrical aspects of the bush theory at an elementary level. Only in the last section, we describe very schematically some mathematical details of this theory.

Let us note in conclusion that the bush theory includes not only a geometrical but also a dynamical part, which the authors hope to consider later in the second part of this review.